\documentclass[11pt]{amsart}
\usepackage{amsmath,amssymb} 
\usepackage{graphicx}
\usepackage[utf8]{inputenc} 
\usepackage[font=small,labelfont=bf]{caption}

\begin{document}

\newtheorem{thm}{Theorem}[section]
\newtheorem{lem}[thm]{Lemma}
\newtheorem{prop}[thm]{Proposition}
\newtheorem{coro}[thm]{Corollary}
\newtheorem{defn}[thm]{Definition}
\newtheorem*{remark}{Remark}

\numberwithin{equation}{section}

\newcommand{\Z}{{\mathbb Z}} 
\newcommand{\Q}{{\mathbb Q}}
\newcommand{\PP}{{\mathbb P}}
\newcommand{\R}{{\mathbb R}}
\newcommand{\C}{{\mathbb C}}
\newcommand{\N}{{\mathbb N}}
\newcommand{\FF}{{\mathbb F}}
\newcommand{\T}{{\mathbb T}}
\newcommand{\fq}{\mathbb{F}_q}

\newcommand{\fixmehidden}[1]{}

\def\scrA{{\mathcal A}}
\def\cB{{\mathcal B}}
\def\Eps{{\mathcal E}}
\def\cI{{\mathcal I}}
\def\scrD{{\mathcal D}}
\def\cF{{\mathcal F}}
\def\cL{{\mathcal L}}
\def\cM{{\mathcal M}}
\def\cN{{\mathcal N}}
\def\cP{{\mathcal P}}
\def\scrR{{\mathcal R}}
\def\scrS{{\mathcal S}}

\newcommand{\rmk}[1]{\footnote{{\bf Comment:} #1}}

\renewcommand{\mod}{\;\operatorname{mod}}
\newcommand{\ord}{\operatorname{ord}}
\newcommand{\TT}{\mathbb{T}}
\renewcommand{\i}{{\mathrm{i}}}
\renewcommand{\d}{{\mathrm{d}}}
\renewcommand{\^}{\widehat}
\newcommand{\HH}{\mathbb H}
\newcommand{\Vol}{\operatorname{vol}}
\newcommand{\area}{\operatorname{area}}
\newcommand{\tr}{\operatorname{tr}}
\newcommand{\norm}{\mathcal N} 
\newcommand{\intinf}{\int_{-\infty}^\infty}
\newcommand{\ave}[1]{\left\langle#1\right\rangle} 
\newcommand{\E}{\mathbb E}
\newcommand{\Var}{\operatorname{Var}}
\newcommand{\Cov}{\operatorname{Cov}}
\newcommand{\Prob}{\operatorname{Prob}}
\newcommand{\sym}{\operatorname{Sym}}
\newcommand{\disc}{\operatorname{disc}}
\newcommand{\CA}{{\mathcal C}_A}
\newcommand{\cond}{\operatorname{cond}} 
\newcommand{\lcm}{\operatorname{lcm}}
\newcommand{\Kl}{\operatorname{Kl}} 
\newcommand{\leg}[2]{\left( \frac{#1}{#2} \right)}  
\newcommand{\id}{\operatorname{id}}
\newcommand{\beq}{\begin{equation}}
\newcommand{\eeq}{\end{equation}}
\newcommand{\bsp}{\begin{split}}
\newcommand{\esp}{\end{split}}
\newcommand{\bra}{\left\langle}
\newcommand{\ket}{\right\rangle}
\newcommand{\diam}{\operatorname{diam}}
\newcommand{\supp}{\operatorname{supp}}
\newcommand{\dist}{\operatorname{dist}}
\newcommand{\sgn}{\operatorname{sgn}}
\newcommand{\inte}{\operatorname{int}}
\newcommand{\Spec}{\operatorname{Spec}}
\newcommand{\sumstar}{\sideset \and^{*} \to \sum}

\newcommand{\LL}{\mathcal L} 
\newcommand{\sumf}{\sum^\flat}
\newcommand{\Hgev}{\mathcal H_{2g+2,q}}
\newcommand{\USp}{\operatorname{USp}}
\newcommand{\conv}{*}
\newcommand{\CF}{c_0} 
\newcommand{\kerp}{\mathcal K}

\newcommand{\gp}{\operatorname{gp}}
\newcommand{\Area}{\operatorname{Area}}

\title[Algebraic delocalization]{Algebraic delocalization for the Schr\"odinger equation on large tori}
\author{Henrik Uebersch\"ar}
\address{Sorbonne Universit\'e, Universit\'e Paris Cit\'e, CNRS, IMJ-PRG, F-75006 Paris, France.}
\email{henrik.ueberschar@cnrs.fr}
\date{\today}

\maketitle
 
\begin{abstract}
Let $\cL$ be a fixed $d$-dimensional lattice. We study the localization properties of solutions of the stationary Schr\"odinger equation with a positive $L^\infty$ potential on tori $\R^d/L\cL$ in the limit, as $L\to\infty$, for dimension $d\leq3$. 

We show that the probability measures associated with $L^2$-normalized solutions, with eigenvalue $E$ near the bottom of the spectrum, satisfy an algebraic delocalization theorem which states that these probability measures cannot be localized inside a ball of radius $r=o(E^{-1/4+\epsilon})$, unless localization occurs with a sufficiently slow algebraic decay.

In particular, we apply our result to Schr\"odinger operators modeling disordered systems, such as the $d$-dimensional continuous Anderson-Bernoulli model, where almost sure exponential localization of eigenfunctions, in the limit as $E\to 0$, was proved by Bourgain-Kenig in dimension $d\geq 2$, and show that our theorem implies an algebraic blow-up of localization length in this limit.
\end{abstract}

\section{Introduction}

We are motivated by the study of the localization properties of eigenfunctions of Schr\"odinger operators which describe disordered systems. For instance, Bourgain-Kenig \cite{BK} considered the Anderson-Bernoulli model on $\R^d$, $d\geq 2$, (the case $d=1$ had previously been treated in \cite{DSS})
\beq
H_{AB}=-\Delta+\sum_{\xi\in\Z^d}\alpha_\xi\varphi(\cdot-\xi)
\eeq
where $\varphi\in C^\infty_c(\R^d)$, $0\leq\varphi\leq1$, $\supp\varphi\subset B(0,1/10)$ with i.i.d. Bernoulli couplings $\alpha_\xi\in\{0,1\}$. One notes that $\inf\sigma(H_\alpha)=0$ almost surely.

Bourgain-Kenig found that solutions of the stationary Schr\"odinger equation $H_{AB}\Psi=E\Psi$ were almost surely exponentially localized in the limit $E\to 0$, therefore proving Anderson localization \cite{A} in the spectral sense. A natural way to study this phenomenon is to consider the operator restricted to a box $[0,L]^d\subset\R^d$, $d\geq 2$, with appropriate boundary conditions and investigate the localization properties of the eigenfunctions in the thermodynamic limit, as $L\to +\infty$.

In this article, instead of considering random operators we will study deterministic operators for any given configuration of scatterer positions and coupling constants. For example, in the case of the continuous Anderson-Bernoulli model, fix a function $\alpha:\Z^d\to\{0,1\}$ and associate with it the Schr\"odinger operator
\beq\label{BK-op}
H_\alpha=-\Delta+\sum_{\xi\in\Z^d}\alpha(\xi)\varphi(\cdot-\xi).
\eeq
We will then study the localization properties of the eigenfunctions of $H_\alpha$ for any given Boolean function $\alpha$ on the lattice $\Z^d$.

A key application of the algebraic delocalization theorem which is proved in this article will be the divergence of the localization length associated with an eigenfunction of $H_\alpha$, as $E\to 0$. We stress for the case $d=2$ that this type of delocalization does not constitute a contradiction to the scaling theory of localization \cite{AALR}: in fact, a similar type of divergence of localization length has been observed in the physics literature for certain models \cite{ERS98,ERS01}. In this context, the authors found multifractal properties of the eigenfunctions, but no full transition to a regime with extended states.

\section{Statement of results} 
We consider Schr\"odinger operators on boxes with periodic boundary conditions (our method should work for other self-adjoint b.c.s and more general domains). We study $L^\infty$ potentials which remain bounded in the thermodynamic limit, with $\inf V=0$ and $\sup V=V_1>0$, a condition which is, in particular, satisfied by the Anderson-Bernoulli operator \eqref{BK-op} for any configuration of coupling constants.

\subsection{Algebraic delocalization} 
Let $d\leq3$ and $\T^d_L=\R^d/L\cL_0$, where $L>0$ is a large parameter, and $\cL_0=\Z(1,0,\cdots,0)\oplus\Z(0,\gamma_1,\cdots,0)\oplus\Z(0,\cdots,0,\gamma_{d-1})$, $\gamma_1,\cdots,\gamma_{d-1}\geq 1$. Let $V\in L^\infty(\T^d_L)$ with $V\geq 0$ and $\inf V=0$. We consider a solution of the Schr\"odinger equation
\beq\label{Sch}
(-\Delta+V)\Psi=E\Psi, \quad \|\Psi\|_{L^2(\T^d_L)}=1, \quad 0<E<1.
\eeq

We show that the $L^2$-density $d\mu_\Psi=|\Psi|^2 d\mu$ associated with a solution $\Psi$ of \eqref{Sch} can only vary subject to certain algebraic constraints. In particular we choose scales $\ell\asymp E^{-1/4+\eta(1-d/4)}$, $\eta\in(0,1/(4-d))$, and $r=\ell E^{-\eta}$. We show that for any $x_0\in\T^d_L$ the ball $B(x_0,\ell)$ contains either less than half of the $L^2$-mass of $\Psi$: $\mu_\Psi(B(x_0,\ell))<1/2$, or the larger ball $B(x_0,r)$ contains less than $1-E$ of the $L^2$-mass of $\Psi$: $\mu_\Psi(B(x_0,r))<1-E$.

\begin{thm}\label{alg_deloc}
There exist absolute constants $c_1,c_2>0$ such that the following holds. Let $\eta\in(0,1/(d-4))$. Denote
$$c_V=c_1(1+\|V\|_\infty^{1/2})^{-1/2}.$$

Let $\Psi$ be a solution of $\eqref{Sch}$ with eigenvalue 
\beq\label{en-int}
E\in[(2c_V/L)^{4/(1+d\eta)},\min(c_V^{4/(1+d\eta)},c_2^{2/d\eta})],
\eeq
where we assume that $L$ is sufficiently large.

Then, we have for any $x_0\in\T^d_L$:\footnote{Note that \eqref{en-int} ensures that 
$1\leq c_V E^{-1/4-d\eta/4}
\leq \tfrac{1}{2}L$}
\beq\label{deloc1}
\int_{B(x_0,c_V E^{-1/4+\eta(1-d/4)})}|\Psi|^2 d\mu < \frac{1}{2} 
\eeq
{\bf or}
\beq\label{deloc2}
\int_{\T^d_L\setminus B(x_0,c_V E^{-1/4-d\eta/4})}|\Psi|^2 d\mu > E.
\eeq 
\end{thm}

\subsection{Blow-up of localization length}
We will illustrate the meaning of Theorem \ref{alg_deloc} by demonstrating that, if we suppose that a solution of the Schr\"odinger equation \eqref{Sch} is localized with respect to a localization center, then our theorem implies that the localization length admits an algebraic singularity, as $E\to 0$, unless the decay of the localization is no faster than algebraic with exponent $d-4$.

Let $\delta:[1,+\infty)\mapsto [0,\tfrac{1}{2}]$ be a continuous, strictly decreasing function.
We say that $\Psi$ is localized with respect to $x_0$ with localization length $\ell_{loc}$ and decay $\delta:[1,+\infty)\mapsto [0,\tfrac{1}{2}]$, if we have for any $r\geq\ell_{loc}$
\beq
\int_{\T^d_L\setminus B(x_0,r)}|\Psi|^2 d\mu \leq \delta\left(\frac{r}{\ell_{loc}}\right).
\eeq

One may now use the result above to deduce lower bounds for the localization length.

Take $\ell_1=c_V E^{-1/4+d\eta/4}$, where $c_V=c_1(1+\|V\|_{L^\infty}^{1/2})^{-1/2}$. Suppose that $\ell_1\geq\ell_{loc}$ (if $\ell_1<\ell_{loc}$, then we have our lower bound). Using the result above we know that for each $x_0\in\T^d_L$ either \eqref{deloc1} or \eqref{deloc2} must hold.
We will show that both inequalities \eqref{deloc1} and \eqref{deloc2} imply a blow-up of localization length, as $E\to 0$.

Inequality \eqref{deloc1} together with the localization hypothesis implies
$$\delta\left(\frac{\ell_1}{\ell_{loc}}\right)\geq \int_{\T^d_L\setminus B(x_0,\ell_1)}|\Psi|^2 d\mu > \frac{1}{2}$$
which is a contradiction, because $\delta(\ell_1/\ell_{loc})\leq \tfrac{1}{2}$. 
So, we have the lower bound $$\ell_{loc}>\ell_1=c_V E^{-1/4+\eta(1-d/4)}.$$

Let us suppose that inequality \eqref{deloc2} holds and deduce a blow-up of localization length in this case. Denote $\ell_2=\ell_1 E^{-\eta}$. We have $\ell_2\geq\ell_1\geq\ell_{loc}$ by our assumption $\ell_1\geq \ell_{loc}$. Therefore, 
$$\delta\left(\frac{\ell_2}{\ell_{loc}}\right)\geq \int_{\T^d_L\setminus B(x_0,\ell_1)}|\Psi|^2 d\mu > E.$$
Because $\delta$ is strictly decreasing, we have
$$\frac{\ell_2}{\ell_{loc}}<\delta^{-1}(E)$$ which yields the lower bound
$$\ell_{loc}>\frac{\ell_2}{\delta^{-1}(E)}=c_V E^{-1/4-d\eta/4}/\delta^{-1}(E).$$

In conclusion, we must have the following lower bound for the localization length
\beq\label{LL_low}
\ell_{loc}>c_V E^{-1/4+\eta(1-d/4)}\min\left(1,\frac{E^{-\eta}}{\delta^{-1}(E)}\right)
\eeq

\subsubsection{Delocalization for exponential decay}
In the case of an exponential decay with respect to a localization center $x_0$, we have for $C,\beta>0$
$$\int_{\T^d_L\setminus B(x_0,r)}|\Psi|^2 d\mu \leq Ce^{-\beta r}=\delta\left(\frac{r}{\ell_{loc}}\right)$$
with $$\delta(r)=Ce^{-\log(2C) r}, \quad \ell_{loc}=\frac{\log(2C)}{\beta}.$$
Since $\delta^{-1}(E)=-\log(E/C)/\log(2C)$, we obtain from \eqref{LL_low} the lower bound (recall $\eta\in(0,1/(4-d))$)
$$\ell_{loc}>c_V E^{-1/4+\eta(1-d/4)}\min\left(1,\log(2C)\frac{E^{-\eta}}{\log(C/E)}\right)$$ 

\subsubsection{Delocalization for sufficiently fast algebraic decay.}
In particular, we see that for any sufficiently fast algebraic decay we may deduce $\ell_{loc}\nearrow+\infty$ as $E\to 0^+$. To see this take $\delta(r)=Cr^{-\alpha}$, which gives
$$\ell_{loc}>c_V E^{-1/4+\eta(1-d/4)}\min\left(1,C^{-1/\alpha}E^{-\eta+1/\alpha}\right).$$

So, for $\alpha\geq 1/\eta$, we have
$$\ell_{loc}\lesssim E^{-1/4+\eta(1-d/4)},$$ because for small $E$, $E^{-\eta+1/\alpha}$ is large.
We then have a blow-up, since $\eta\in(0,1/(4-d))$. 

Moreover, if $\alpha<1/\eta$, then $E^{-\eta+1/\alpha}$ is small if $E\to 0$. So, the lower bound is of order 
$$\ell_{loc}\lesssim E^{-1/4-d\eta/4+1/\alpha},$$ and we have a blow-up provided $1/\alpha<1/4+d\eta/4$. This yields the general condition $\alpha>4/(1+d\eta)$ to ensure a blow-up of localization length. Since $\eta\in(0,1/(4-d))$, we can rule out any decay faster than exponent $4-d$.

\subsection{Generalization to spectral projectors}

In the context of Anderson localization it is natural to study the localization properties of all eigenfunctions in a given spectral window $\cI=[E,2E]$, $0<E<1$. Let $h:\R\to\R$ be a continuous, compactly supported function with $\supp h\subset\cI$. 

We introduce
$$\Pi_h(x,y)=\sum_{\lambda}h(\lambda)\Psi_\lambda(x)\overline{\Psi_\lambda}(y).$$

We note that, while individual eigenfunctions are expected to be exponentially localized with respect to a localization center, these centers may be different for each eigenfunction. Therefore one considers a spectral projector $\Pi_h(x,y)$ which ought to satisfy an exponential decay with respect to the distance $|x-y|$.

Let us fix $x_0\in\T^d_L$ and let $F=\Pi_h(\cdot,x_0)/\|\Pi_h(\cdot,x_0)\|_{L^2(\T^d_L)}$. Let $\mu_h$ be the probability measure associated with the probability density $d\mu_h=|F|^2 d\mu$. We have the following algebraic delocalization theorem which is a generalization of Theorem \ref{alg_deloc}.
\begin{thm}\label{alg_deloc_2}
Assume that $E$ satisifies the conditions of Theorem \ref{alg_deloc}. Let $B_\ell\subset\T^d_L$ be any ball of radius $\ell=c_V E^{-1/4+\eta(1-d/4)}$, and $B_r\subset \T^d_L$ be a ball with the same center and radius $r=\ell E^{-\eta}$. We have either $\mu_h(B_\ell)<\tfrac{1}{2}$ or $\mu_h(B_r)<1-E$.
\end{thm}

Arguing as in the previous section one may now deduce a blow-up of localization length for any decay faster than exponent $d-4$.

\section{Proof of Theorem \ref{alg_deloc}}

We argue by contradiction. Recall $\ell=c_1E^{-1/4+\eta(1-d/4)}(1+\|V\|_{L^\infty}^{1/2})^{-1/2}$. Let us suppose there exists $x_0\in\T^d_L$ such that
\beq\label{loc1}
\int_{B(x_0,\ell)}|\Psi|^2 d\mu \geq \frac{1}{2} 
\eeq
and 
\beq\label{loc2}
\int_{\T^d_L\setminus B(x_0,\ell E^{-\eta})}|\Psi|^2 d\mu \leq E.
\eeq

\subsection{A variation bound}

Let $r=\ell E^{-\eta}$ for $\eta\in(0,1/(4-d))$. We recall that our assumption \eqref{en-int} implies $1\leq r\leq \tfrac{1}{2}L$.
Let $\Lambda_{2r}\subset\T^d_L$ be a box of side length $2r$ centered on $x_0$ inside the torus such that 
$B(x_0,r)\subset \Lambda_{2r}$. 
Let $\chi\in C^{\infty}(\T^d_L)$ be a smoothed indicator in the sense that $\Lambda_{2r}\subset\supp\chi\subset \Lambda_{2r+1}$, and  $\chi|_{\Lambda_{2r}}=1$. Moreover we may choose $\chi$ in such a way that $\sup|\nabla\chi|\leq k_1$ and $\sup|\Delta\chi|\leq k_2$ for absolute constants $k_1, k_2$. 

Let $x_1,x_2\in \Lambda_{2r}$. Let $r'=2r+1$. We recall that the kernel $G_\lambda$ of the resolvent $(-\Delta-\lambda)^{-1}$ on $\Lambda_{r'}$ can be expanded with respect to the o. n. b. of exponentials $e(\xi\cdot x/r')r'^{-d/2}$, $\xi\in\Z^d$, as
$$G_\lambda(x,y)=-\frac{1}{r'^d\lambda}+\sum_{\xi\in\Z^2\setminus\{0\}}\frac{1}{4\pi^2|\xi/r'|^2-\lambda}\frac{e(\xi\cdot (x-y)/r')}{r'^d},$$
where $\lambda\notin\sigma(-\Delta)$ and we denote $e(\tau)=e^{2\pi i \tau}$.

We then have
$$G_\lambda(x,x_1)-G_\lambda(x,x_2)=\sum_{\xi\in\Z^2\setminus\{0\}}\frac{e(\xi\cdot (-x_1)/r')-e(\xi\cdot (-x_2)/r')}{4\pi^2|\xi/r'|^2-\lambda}\frac{e(\xi\cdot x/r')}{r'^d}.$$
We note that we may pass to the limit $\lambda\to 0$ and obtain an $L^2$ function on $\Lambda_{r'}$.

Let us define $b\in L^2(\Lambda_{r'})$ by $$b(x):=\frac{1}{4\pi^2}\sum_{\xi\in\Z^2\setminus\{0\}}\frac{e(-\xi\cdot x_1/r')-e(-\xi\cdot x_2/r')}{r'^{d/2}|\xi/r'|^2}\frac{e(\xi\cdot x/r')}{r'^{d/2}}$$
so that $$-\Delta b=\delta_{x_1}-\delta_{x_2}.$$

Recall $x_1, x_2\in \Lambda_{2r}$. By construction of the smooth cut-off $\chi$ above, we have $\chi\Psi\in C^2(\Lambda_{r'})$ and $\chi\Psi|_{\partial \Lambda_{r'}}$=0, which ensures that the cutoff $\chi\Psi$ satisfies periodic boundary conditions. We have the identity 
\beq
\Psi(x_1)-\Psi(x_2)=
\int_{\Lambda_{r'}}(-\Delta b)\,\chi\Psi \,d\mu=\int_{\Lambda_{r'}} b\,(-\Delta(\chi \Psi)) \,d\mu
\eeq

We may now expand $$-\Delta(\chi \Psi)=-\Psi \Delta\chi-2\nabla\chi\cdot\nabla\Psi-\chi\Delta\Psi.$$

We estimate the integral term by term. We start with the third term
\beq
\int_{\Lambda_{r'}} b \,\chi \,(-\Delta\Psi) \,d\mu
=E\int_{\Lambda_{r'}} b \,\chi \,\Psi \,d\mu-\int_{\Lambda_{r'}} b \,\chi \,V\Psi \,d\mu 
\eeq
where we used $(-\Delta+V)\Psi=E\Psi$.

Moreover, the identity
$$0\leq\int_{\T^d_L}|\nabla \Psi|^2 d\mu=\int_{\T^d_L}-\Delta\Psi \cdot \overline{\Psi} d\mu
=\int_{\T^d_L}(E-V)|\Psi|^2 d\mu$$
yields $$\int_{\Lambda_{r'}}V|\Psi|^2 d\mu\leq\int_{\T^d_L}V|\Psi|^2 d\mu\leq E,$$ where we used $\|\Psi\|_{L^2(\T^d_L)}=1$.

Hence, Cauchy-Schwarz yields
\beq
\begin{split}
\Big|\int_{\Lambda_{r'}} b \,\chi \,V\Psi \,d\mu\Big| 
&\leq \|b\chi\|_{L^2(\Lambda_{r'})}\|V\|_{L^\infty}^{1/2}\left(\int_{\T^d_L}V|\Psi|^2\right)^{1/2}\\
&\leq \|b\|_{L^2(\Lambda_{r'})}\|V\|_{L^\infty}^{1/2} E^{1/2}
\end{split}
\eeq
(note $\|b\chi\|_{L^2(\Lambda_{r'})}\leq\|b\|_{L^2(\Lambda_{r'})}$) 
and 
\beq
\begin{split}
\|b\|_{L^2(\Lambda_{r'})} 
\leq \frac{1}{2\pi^2}\left(\sum_{\xi\in\Z^2\setminus\{0\}}\frac{1}{r'^d|\xi/r'|^4}\right)^{1/2}
=\frac{r'^{(4-d)/2}L_d(2)^{1/2}}{2\pi^2}
\end{split}
\eeq
 where we denote the Dirichlet series
$$L_d(s)=\sum_{n=1}^\infty r_d(n)n^{-s}$$ associated with $r_d(n)$, the number of ways the integer $n$ can be represented as a number of $d$ squares. Note that the series converges for $s=2$, since $d\leq 3$.
 
Moreover,   
$$E\Big|\int_{\Lambda_{r'}}b\chi\Psi d\mu\Big|\leq E\|b\chi\|_{L^2(\Lambda_{r'})}\leq \frac{Er'^{(d-4)/2}L_d(2)^{1/2}}{2\pi^2}$$
where we used $\|\Psi\|_{L^2(\T^d_L)}=1$ and $\Lambda_{r'}\subset\T^d_L$.

Let us continue with the second term. We have
\beq
\Big|\int_{\Lambda_{r'}}b\nabla\chi\cdot\nabla\Psi d\mu\Big|
\leq \|b\|_{L^2(\Lambda_{r'})}\|\nabla\chi\cdot\nabla\Psi\|_{L^2(\Lambda_{r'})}
\eeq
and
\begin{align*}
\|\nabla\chi\cdot\nabla\Psi\|_{L^2(\Lambda_{r'})}
\leq \left(\int_{\Lambda_{r'}}|\nabla\chi|^2|\nabla\Psi|^2d\mu\right)^{1/2}\\
\leq k_1\||\nabla\Psi|\|_{L^2(\T^d_L)}
\end{align*}
where we recall $\sup|\nabla\chi|\leq k_1$, and, as we saw above,
$$\||\nabla\Psi|\|_{L^2(\T^d_L)}^2=\int_{\T^d_L}(E-V)|\Psi|^2d\mu\leq E,$$
where we used $\|\Psi\|_{L^2(\T^d_L)}=1$ and $V\geq 0$.

So, in summary, we bound the second term as follows:
$$\Big|\int_{\Lambda_{r'}}b\nabla\chi\cdot\nabla\Psi d\mu\Big|\leq k_1\frac{L_d(2)^{1/2}}{2\pi^2}r'^{(4-d)/2}E^{1/2}.$$

For the first term, we use \eqref{loc2}:
\beq
\begin{split}
\Big|\int_{\Lambda_{r'}}b \,\Psi \,\Delta\chi \,d\mu\Big|
\leq &\|b\|_{L^2(\Lambda_{r'})}\sup|\Delta\chi|\left(\int_{\supp(\Delta\chi)}|\Psi|^2 d\mu\right)^{1/2}\\
\leq &k_2 \frac{L_d(2)^{1/2}}{2\pi^2}r'^{(d-4)/2} E^{1/2}
\end{split}
\eeq
where we recall $\sup|\Delta\chi|\leq k_2$, as well as $\|b\|_{L^2(\Lambda_{r'})}\leq r'^{(4-d)/2}L_D(2)^{1/2}/2\pi^2$.

In the estimate of the integral over $\supp(\Delta\chi)$, we note that by construction of the cutoff function $\chi$ we have $\supp(\Delta\chi) \subset \Lambda_{r'}\setminus\Lambda_{2r}$ (recall $r'=2r+1)$ and, therefore,
$$\supp(\Delta\chi)\subset \T^d_L\setminus \Lambda_{2r} \subset \T^d_L\setminus B(x_0,r)$$
and, thus, using \eqref{loc2} (recall $r=\ell E^{-\eta}$), we obtain
$$\int_{\supp(\Delta\chi)}|\Psi|^2 d\mu \leq \int_{\T^d_L\setminus B(x_0,r)}|\Psi|^2 d\mu \leq E.$$

Combining all of the above estimates, we obtain
\beq
\Big|\Psi(x_1)-\Psi(x_2)\Big|
\leq c r^{(4-d)/2}(1+\|V\|_{L^\infty}^{1/2})E^{1/2}
\eeq
where $c>0$ denotes an absolute constant and $r'=2r+1\leq 3r$, as $r\geq1$.

\subsection{Proof}

In order to prove the result we will first of all show that our assumption 
$$\int_{B(x_0,\ell)}|\Psi|^2 d\mu\geq\frac{1}{2}$$
implies that there exists $x'\in B(x_0,\ell)$ s. t. $|\Psi(x')|\geq \sqrt{\frac{1}{2\pi}}\ell^{-d/2}$.

To see this, we argue by contradiction. Suppose we have $|\Psi(x)|< \sqrt{\frac{1}{2\pi}}\ell^{-d/2}$ for all $x\in B$.
Then, $$\int_B|\Psi|^2d\mu<\frac{1}{2\pi}\ell^{-d}\Vol(B)=\frac{1}{2}.$$
  
For any $x\in \Lambda_{2r}$ we have 
\beq
\begin{split}
|\Psi(x)|
&\geq |\Psi(x')|-|\Psi(x)-\Psi(x')|\\
&\geq \frac{1}{\sqrt{2\pi}}\ell^{-d/2}-c(1+\|V\|_{L^\infty}^{1/2})r^{(4-d)/2}E^{1/2}.
\end{split}
\eeq

In order to have a lower bound of order $\ell^{-d/2}$, we then need
$$c'\ell^{-d/2}\geq (1+\|V\|_{L^\infty}^{1/2})r^{(4-d)/2}E^{1/2}=(1+\|V\|_{L^\infty}^{1/2})\ell^{(4-d)/2} E^{1/2-\eta(4-d)/2}$$
for some absolute constant $c'>0$, which, in turn, is equivalent to
$$\ell\leq \left(\frac{c'}{1+\|V\|_{L^\infty}^{1/2}}\right)^{1/2}E^{-1/4+\eta(1-d/4)}$$
and we recall that this holds with $c_1=c'^{1/2}$.

Then we have for any $x\in \Lambda_{2r}$ 
$$|\Psi(x)|\geq c''\ell^{-d/2},$$
for an absolute constant $c''>0$.
This then yields
$$\int_{\Lambda_{2r}}|\Psi(x)|^2d\mu \geq c''^2\frac{(2r)^d}{\ell^d}=2^d c''^2 E^{-\eta d}$$
and, thus, for an absolute constant $0<c_2<c''$ we have $E\leq c_2^{2/d\eta}<(2c''^{2/d})^{1/\eta}$ and we arrive at a contradiction to the normalization $\|\Psi\|_{L^2(\T^d_L)}=1$.

\section{Proof of Theorem \ref{alg_deloc_2}}

For given $x_0\in\T^d_L$, we introduce
$$F=\frac{\sum_{\lambda\in[E,2E]}h(\lambda)\overline{\Psi_\lambda}(x_0)\Psi_\lambda}{(\sum_{\lambda\in[E,2E]}|\Psi_\lambda(x_0)|^2)^{1/2}}.$$
We sketch the argument which is very similar to the one above for individual eigenfunctions.
In particular, we have $-\Delta F=\tilde{F}-VF$,
where
$$\tilde{F} = \frac{\sum_{\lambda\in[E,2E]}\lambda h(\lambda)\overline{\Psi_\lambda}(x_0)\Psi_\lambda}
{(\sum_{\lambda\in[E,2E]}|\Psi_\lambda(x_0)|^2)^{1/2}}$$
and $$\|\tilde{F}\|_{L^2}^2=\frac{\sum_{\lambda\in[E,2E]}\lambda^2 h(\lambda)^2|\Psi_\lambda(x_0)|^2}
{\sum_{\lambda\in[E,2E]}|\Psi_\lambda(x_0)|^2}\leq 4\|h\|_{L^{\infty}}^2E^2$$

Following the argument above we get for $x_1,x_2\in\Lambda_{2r}$
$$|F(x_1)-F(x_2)|\leq \Big| \int_{\T^d_L}b \, \Delta(\chi F) \,d\mu \Big|$$

Let us estimate the integral with the Laplace term
$$\int_{\T^d_L}b \,\chi \,\Delta F \,d\mu
=\int_{\T^d_L}b \,\chi \tilde{F} \,d\mu - \int_{\T^d_L}b \,\chi VF \,d\mu$$

For the first term, we find an estimate of the same order as above
$$\Big|\int_{\T^d_L}b \,\chi \tilde{F} \,d\mu\Big| \leq 2E\|b\|_2\lesssim r'^{(4-d)/2}E^{1/2}.$$

For the second term, we have
$$\Big|\int_{\T^d_L}b \,\chi VF \,d\mu\Big| \leq \|b\|_{L^2(\Lambda_{r'})}\|V\|_{L^\infty}\left(\int_{\T^d_L}V|F|^2d\mu\right)^{1/2}$$
and 
$$0\leq \int_{\T^d_L}|\nabla F|^2 d\mu
=\int_{\T^d_L}(-\Delta F) F d\mu
=\int_{\T^d_L}\tilde{F} F d\mu-\int_{\T^d_L}V |F|^2 d\mu$$
which yields
$$\int_{\T^d_L}V|F|^2d\mu\leq \int_{\T^d_L}\tilde{F} F d\mu \leq \|\tilde{F}\|_2\|F\|_2\leq 2\|h\|_{L^\infty}E$$
And again, we get an estimate of the same order as above
$$\Big|\int_{\T^d_L}b \,\chi VF \,d\mu\Big| \lesssim \|V\|_{L^\infty}r'^{(4-d)/2}E^{1/2}$$.

Moreover, for the gradient term we get the same estimate as above due to
$$\int_{\T^d_L}|\nabla F|^2 d\mu
=\int_{\T^d_L}\tilde{F} F d\mu-\int_{\T^d_L}V |F|^2 d\mu\leq \int_{\T^d_L}\tilde{F} F d\mu\leq 2\|h\|_{L^\infty}E.$$

\end{document}